\documentclass[twocolumn,showpacs,preprintnumbers,amsmath,amssymb]{revtex4}
\usepackage[dvips]{graphicx}
\usepackage[tight]{subfigure}
\usepackage{subfigure}

\begin{document}

\author{T. Kroll$^1$}
\email{[]t.kroll@ifw-dresden.de}
\author{A.A. Aligia$^{2}$}
\author{G.A. Sawatzky$^{3}$}
\affiliation{$^{1}$IFW Dresden, P.O. Box 270016, D-01171 Dresden, Germany}
\affiliation{$^{2}$ Comisi\'on Nacional de Energ\'ia At\'omica, Centro At\'omico
Bariloche and Instituto Balseiro, 8400 S. C. de Bariloche, Argentina}
\affiliation{$^{3}$ Department of Physics and Astronomy, University of British Columbia,
6224 Agricultural Road, Vancouver, British Columbia, Canada V6T 1Z1}

\title{Polarization dependence of x-ray absorption spectra in Na$_x$CoO$_2$}

\begin{abstract}
In order to shed light on the electronic structure of Na$_x$CoO$_2$, and
motivated by recent Co L-edge X-ray absorption spectra (XAS) experiments
with polarized light, we calculate the electronic spectrum of a CoO$_6$
cluster including all interactions between 3d orbitals. We obtain the ground
state for two electronic occupations in the cluster that correspond
nominally to all O in the O$^{-2}$ oxidation state, and Co$^{+3}$ or Co$%
^{+4} $. Then, all excited states obtained by promotion of a Co 2p
electron to a 3d electron, and the corresponding matrix elements
are calculated. A fit of the observed experimental spectra is good
and points out a large Co-O covalency and cubic crystal field
effects, that result in low spin Co 3d configurations. Our results
indicate that
% although the distortion of the cubic shape of the octahedra is important,
the effective hopping between different Co atoms plays a major
role in determining the symmetry of the ground state in the
lattice. Remaining quantitative discrepancies with the XAS
experiments are expected to come from composition effects of
itineracy in the ground and excited states.

% PACS numbers: 71.27.+a, 71.70.–d, 74.70.–b, 78.70.Dm
\end{abstract}

\maketitle

\section{Introduction}

In recent years, there has been
intense research  on doped cobaltates
because of their interesting transport, magnetic and superconducting
properties. $\mathrm{Na}_{x}\mathrm{CoO}_{2}$ shows an exceptionally high
thermopower over the doping range $0.5\le x\le 0.9$, while also displaying
low resistivity and low thermal conductivity \cite%
{Terasaki_PRB02,Mikami_JJAP03}. These properties are those wished in
materials with potential technological applications in refrigeration. In
2003 Takada and coworkers found a superconducting state in $\mathrm{%
Na_{0.3}CoO2\cdot 1.3H_{2}O}$ at temperatures lower than 5 K \cite%
{Takada_Nature03}. The system $\mathrm{Na}_{x}\mathrm{CoO}_{2}$ is also
interesting because of its similarities to the high-$T_{c}$ copper oxide
superconductors, since both systems contain alternating layers consisting of
oxygen and spin-1/2 transition metal ions separated by layers of lower
conductivity in an anisotropic crystal structure. $\mathrm{Na}_{x}\mathrm{CoO%
}_{2}$ consists of alternating layers of $\mathrm{CoO_{2}}$ and sodium, and
becomes superconducting only when hydrated with water ($\mathrm{%
Na_{0.35}CoO_{2} \cdot 1.3H_{2}O}$)\cite{Takada_Nature03}. However, in
contrast to most cuprates, the $\mathrm{CoO_{2}}$ layers consist of
edge-sharing $\mathrm{CoO_{6}}$ octahedra, arranged in such a way that the
cobalt and oxygen ions form a triangular lattice. The superconducting
resonance-valence-bond state for which some indications exist in the square
lattice of the cuprates \cite{and,bati}, is believed to be favored in a
triangular lattice \cite{Baskaran_PRL03}.

In the most simple picture, the effect of sodium doping is only to modify
the nominal ratio of $\mathrm{Co^{3+}}$ to $\mathrm{Co^{4+}}$ depending on
the sodium content. This in turn, as in the cuprates, modifies the number of
free carriers and affects dramatically the conducting and magnetic
properties of the system. $x = 0$ corresponds to the highly correlated
(charge transfer or Mott) insulating limit. Magnetic susceptibility
measurements \cite{Mikami_JJAP03,Bayrakci_PRB04} have shown evidence of
antiferromagnetic long-range order below 20 K in a doping range of $\mathrm{%
0.75\le x\le 0.9.}$ From anisotropic dc magnetic susceptibility and $\mu $SR
measurements on a $\mathrm{Na_{0.82}CoO_{2}}$ crystal one concludes that the
Co spin is oriented along the crystal $c$ axis \cite{Bayrakci_PRB04}.

Detailed knowledge of the electronic structure of this system is of course
important for the understanding of all its properties. Ideally, from the
knowledge of important parameters such as the $d-d$ Coulomb interaction
which includes exchange, the different hopping terms, and the
charge-transfer energy one directly gains information about the band gap and
the character of the electronic structure as it is explained in the
classification scheme done by Zaanen, Sawatzky, and Allen \cite{Zaanen_PRL85}%
. Additionally, knowing important orbitals and interactions in an
intermediate energy scale accessible in optical experiments, like the
three--band Hubbard model in the case of cuprates \cite{hyb}, one would
derive an effective low-energy model (like the $t-J$ model based on
Zhang-Rice singlets \cite{zhang, Eskes_PRL88, simon, feiner, sim2} for the
cuprates) which describes the relevant low-energy physics. As an example,
interesting phenomena have been explained by the formation of the Zhang-Rice
singlets, such as the unusual magnetic and transport properties in $\mathrm{%
(La,Ca)_{x}Sr_{14-x}Cu_{24}O_{41}}$ \cite{Kroll_JMM05} and other cuprates.
Similar low-energy reduction procedures have been followed for the
nickelates \cite{Batista_EPL98} and magnetic double perovskites \cite{paula}%
. Instead, a systematic low-energy reduction has not been carried out for $%
\mathrm{Na}_{x}\mathrm{CoO}_{2}$ and it seems that no consensus exists on
the low-energy effective theory, in spite of the experimental and
theoretical effort concerning the electronic structure of the system. At
energy scales of 1 eV or larger, valuable information on this electronic
structure is provided by local spectroscopic probes, like x-ray absorption
spectroscopy (XAS) \cite{Wu_PRL05, Kroll_UJS10}, and x-ray photoemission
(XPS) \cite{Chainani_PRB04}, while angle resolved photoemission (ARPES) \cite%
{Hasan_PRL04, Yang_PRL04, yang2} brings information on the
dispersion relation just below the Fermi energy. In particular
XAS, including a recent study of its dependence on the
polarization of the incident light \cite{Wu_PRL05, Kroll_UJS10},
indicate that $\mathrm{Na}_{x}\mathrm{CoO}_{2}$ exhibits the
character of a doped charge-transfer insulator
\cite{Zaanen_PRL85}, rather than a doped Mott--Hubbard insulator,
and the ground state contains a significant O 2p portion. Several
calculations of the band structure use a single band effective
model \cite{Baskaran_PRL03, Singh_PRB00, Singh_PRB03,
Rosner_BJP03, wang}. These descriptions can be justified if the
lattice distortion or crystal field break the degeneracy of the 3d
$t_{2g}$ orbitals which are otherwise degenerate in a cubic
environment, and if the O degrees of freedom can be integrated out
in a similar way as in the cuprates. However, Koshibae and Maekawa
based on an estimation of the effects of the above mentioned
lattice distortion, and simple geometrical arguments, proposed an
alternative scenario, with four interpenetrating Kagom\'{e}
sublattices hidden in the $\mathrm{CoO_{2}}$ layer
\cite{Koshibae_PRL03}. This scenario was further supported by
Indergand \textit{et al.} who derived the different hopping
elements and interactions in the ensuing multi-band
model \cite{Indergand_PRB05}. Starting from this picture, Khalliulin \textit{%
et al.} argued that the spin-orbit coupling of the correlated electrons on
the $t_{2g}$ level, although relatively small ($\sim 80$ meV \cite{blazey})
strongly influences the coherent part of the wave functions and the Fermi
surface topology at low doping and favors triplet superconductivity, in
marked contrast to the cuprates \cite{Khalliulin_PRL04}.

In this work, we solve exactly a cluster composed of a Co atom and its six
nearest-neighbor O atoms for the geometry and oxidations states relevant to $%
\mathrm{Na}_{x}\mathrm{CoO}_{2}$. The aim is to learn more about the
electronic structure of the system in an intermediate energy scale by
comparison with experimental results on the polarization dependence of XAS
spectra \cite{Kroll_UJS10}. We confirm previous results concerning the
highly covalent Co-O bond and the low spin state of Co \cite{Wu_PRL05,
Kroll_UJS10}. In addition an analysis of the dependence of the XAS
intensities
on the polarization direction of the incident light,
support the scenario of Koshibae and Maekawa that the
effective Co-Co hopping through the intermediate O
sites is an essential part of the low energy physics.
Additionally it will be shown, that due to the strong covalency
a large amount of holes
reside at the oxygen sites.

The formalism and relevant equations are presented in Section II. The
results and its comparison to experiment are included in Section III.
Section IV contains a summary and discussion.

\section{Method}

\subsection{The model}

An adequate starting point to describe the electronic structure of $\mathrm{%
Na}_{x}\mathrm{CoO}_{2}$ is a multiband Hubbard model including all Co 3d
orbitals and O 2p orbitals, and their interactions (like the three-band
Hubbard model for the cuprates \cite{hyb}). This model cannot be solved
exactly and one has to resort to some approximation. Calculations starting
from first principles approximate the interactions and are known to lead to
wrong results when the on-site Coulomb repulsion $U_{d}$ between
transition-metal 3d electrons is important. For example, they predict that
LaCuO$_{4}$ is a paramagnetic metal, while experimentally it is an
antiferromagnetic insulator. In this work we take the opposite approach and
retain exactly all 3d-3d interactions at the price to consider only one Co
atom and its six nearest-neighbor O atoms, with octahedral $O_{h}$ symmetry.
The effects of the distortions are discussed in section III. We also make
some additional approximations which are not essential but reduce the size
of the matrices of the Hamiltonian in each symmetry sector. This allows us
to have a calculational tool flexible enough to allow a fit of the XAS
spectra in a reasonable amount of computer time. For example, we neglect the
interactions between O 2p holes and this allows us to take us advantage of a
restricted basis for these orbitals which has the same symmetry as the 3d $%
t_{2g}$ and $e_{g}$ orbitals. The other O 2p orbitals become irrelevant in
the absence of O-O interactions. For the description of the Co L-edge XAS
spectra, we also need to include in the Hamiltonian the energy of the Co 2p
core hole and its repulsion with the Co 3d holes \cite{vanderLaan_PRB81,
Kakehashi_PRB84, hyb}.

The Hamiltonian that describes the dynamics of holes inside the $\mathrm{CoO}%
_{6}$ cluster is

\begin{eqnarray}
H &=&\sum_{\alpha \in t_{2g},\sigma }\epsilon _{t_{2g}}d_{\alpha \sigma
}^{+}d_{\alpha \sigma }+\sum_{\alpha \in e_{g},\sigma }\epsilon
_{e_{g}}d_{\alpha \sigma }^{+}d_{\alpha \sigma }+\sum_{jm}\epsilon
_{j}c_{jm}^{+}c_{jm}  \nonumber \\
&&+\sum_{i\alpha \sigma }\epsilon _{\mathrm{O}}p_{i\alpha \sigma
}^{+}p_{i\alpha \sigma }+\sum_{i\neq j\alpha \beta \sigma }t_{i}^{\alpha
\beta }(p_{i\alpha \sigma }^{+}d_{\beta \sigma }+\mathrm{H.c.})  \nonumber \\
&&+\sum_{i\neq k\alpha \beta \sigma }\tau _{ij}^{\alpha \beta }p_{i\alpha
\sigma }^{+}p_{k\beta \sigma }+\frac{1}{2}\sum_{\lambda \mu \nu \rho
}V_{\lambda \mu \nu \rho }d_{\lambda }^{+}d_{\mu }^{+}d_{\rho }d_{\nu }
\nonumber \\
&&+U_{cd}\sum_{jm\alpha \sigma }c_{jm}^{+}c_{jm}d_{\alpha \sigma
}^{+}d_{\alpha \sigma }\text{. }  \label{Hamiltonian}
\end{eqnarray}
Here $p_{i\alpha \sigma }^{+}$ creates a hole on the O 2p orbital $\alpha $
at site $i$ with spin $\sigma $. The operator $d_{\alpha \sigma }^{+}$ has
an analogous meaning for the Co 3d orbitals. Similarly $c_{jm}^{+}$ creates
a Co 2p core hole with angular momentum $j$ and projection $m$.

The first two terms of Eq. (\ref{Hamiltonian}) describe the energy of the $%
t_{2g}$ and $e_{g}$ Co 3d orbitals split by a crystal field $\epsilon
_{t_{2g}}-\epsilon _{e_{g}}=10Dq$. We note that this difference can be
called the ``ionic'' contribution to the crystal field splitting, since
after hybridization with the O orbitals, the difference between mixed
$t_{2g} $ and $e_{g}$ orbitals is much larger
due to what is called the ligand field contributions.
The third term describes the
energy of the Co core hole, with a splitting $\epsilon _{3/2}-\epsilon
_{1/2}\sim 15 $ eV. The fourth term corresponds to the energy of the O 2p
orbitals. The next two terms represent the Co-O and O-O hopping,
parametrized as usual, in terms of the Slater-Koster parameters \cite{slat}.
The term before the last includes all interactions between 3d orbitals
originated by the Coulomb repulsion of electrons in the 3d shell. The last
term describes the repulsion between the Co 3d holes and a Co 2p core hole
if present. We have neglected the
exchange and  higher multipole Coulomb interactions  between these
orbitals. Their effects will be discussed later.
Spin-orbit coupling of the 3d electrons is
negligible compared with the other energies in the problem ($\sim 80$ meV
\cite{blazey}) and was also neglected.

The matrix elements of the interactions inside the 3d shell are given by
\cite{nege}

\begin{equation}
V_{\lambda \mu \nu \rho }=\int d\mathbf{r}_{1}d\mathbf{r}_{2}\bar{\varphi}%
_{\lambda }(\mathbf{r}_{1})\bar{\varphi}_{\mu }(\mathbf{r}_{2})\frac{e^{2}}{|%
\mathbf{r}_{1}-\mathbf{r}_{2}|}\varphi _{\nu }(\mathbf{r}_{1})\varphi _{\rho
}(\mathbf{r}_{2}),  \label{integ}
\end{equation}
where $\varphi _{\lambda }(\mathbf{r}_{1})$ is the wave function of the
spin-orbital $\lambda $. We have calculated all these integrals in terms of
the Slater parameters $F_{0}$, $F_{2}$, and $F_{4}$ using known methods of
atomic physics \cite{cond,bal}. The resulting form of this interaction term
is long and we do not reproduce it here. The terms including only $t_{2g}$
orbitals are described in Ref \cite{ruo}, and those among $e_{g}$ ones are
described for example in Ref. \cite{Batista_EPL98}. Some of the remaining
terms are listed in Ref. \cite{bal}. In particular, the value of the Coulomb
repulsion between electrons of opposite spin at the same orbital is $%
U_{d}=F_{0}+4F_{2}+36F_{4}$, and the
spin-spin interaction
between 3d orbitals
of the same irreducible representation is $J_{t_{2g}}=3F_{2}+20F_{4}$, and $%
J_{e_{g}}=4F_{2}+15F_{4}$.

In $\mathrm{Na}_{x}\mathrm{CoO}_{2}$, the formal oxidation state of the O
ions is -2, and that of Co depends on the sodium content $x$, being $\mathrm{%
Co^{3+}}$ for $x=1$, and $\mathrm{Co^{4+}}$ for $x=0$. This means that our
cluster has to be solved for a total of four holes (configurations $%
d^{6}L^{0}$, $d^{7}L^{1}$, $d^{8}L^{2}$, etc.) in the former case and five
holes (configurations $d^{5}L^{0}$, $d^{6}L^{1}$, etc.) in the latter. The
respective probabilities are assumed to be $x$ and $1-x$. For simplicity, we
will label these two cases as $\mathrm{Co^{3+}}$ and $\mathrm{Co^{4+}}$
respectively, although as we shall see there is a significant degree of
covalency and the real Co oxidation states are smaller.

\subsection{The x-ray absorption intensity}

The ground state of the cluster in both above mentioned cases (for $\mathrm{%
Co^{3+}}$ and $\mathrm{Co^{4+}}$) does not contain any Co 2p core hole. The
excitation with light promotes a Co 2p electron to the Co 3d shell, or in
other words, creates a core hole and destroys a 3d hole. In the dipolar
approximation, the effect of the light is calculated in time-dependent
perturbation theory from the addition of the term $\mathcal{H}_{L}=\mathbf{p}
\cdot \mathbf{A}$ to the Hamiltonian, where $\mathbf{p}$ is the momentum
operator and $\mathbf{A}=(A_{x},A_{y},A_{z})$ is the vector potential.
Except for an unimportant prefactor, the relevant part of $\mathcal{H}_{L}$
can be deduced from symmetry arguments and general physical considerations:
it should be rotationally invariant, spin independent, and contain terms
that destroy a 3d hole and create a core hole. This leads to the operator
\begin{equation}
\mathcal{H}_{L}=\sum_{l_{z}\sigma }(O_{2l_{z}\sigma }^{+}d_{2l_{z}\sigma }+%
\mathrm{H.c.)}.  \label{hl1}
\end{equation}
Here the destruction operators of the 3d holes $d_{2l_{z}\sigma }$ are
expressed in terms of $l_{z},$ the orbital angular momentum projection. $%
O_{2l_{z}\sigma }^{+}$ is an irreducible operator that transforms like an
angular momentum $L=2$ with projection $l_{z}$, constructed combining the
components of $\mathbf{A}$ in spherical harmonics ($A_{1l_{z}}$) with the
core hole operators $c_{1l_{z}\sigma }^{+}$ that create a Co 2p hole with
angular momentum projection $l_{z}$ and spin $\sigma $. The expression of
the five components of $O_{2l_{z}\sigma }^{+}$ ($l_{z}=-2$ to 2) is derived
easily using Clebsch-Gordan coefficients. Similarly, using these
coefficients, it is easy although lengthy, to express the light operator $%
\mathcal{H}_{L}$ in terms of the basis for the orbitals used in the
Hamiltonian (Eq. (\ref{Hamiltonian})) and Cartesian coordinates of $\mathbf{A%
}$:

\begin{equation}
\mathcal{H}_{L}=\sum_{\beta jm\alpha \sigma }a_{\beta jm\alpha \sigma
}A_{\beta }p_{jm}^{+}d_{\alpha \sigma }+\mathrm{H.c.}  \label{hl2}
\end{equation}
This is the most convenient form for our purposes.

Using Fermi's golden rule and neglecting an unimportant prefactor, the XAS
intensity becomes
\begin{equation}
I=\sum_{if}p_{i}|\langle f|\mathcal{H}_{L}|i\rangle |^{2}\delta
(E_{f}-E_{i}-\hbar \omega ).  \label{ixas}
\end{equation}
Here $|i\rangle $, is one of the two possible initial states (ground state
for $\mathrm{Co^{3+}}$ or $\mathrm{Co^{4+}}$), $p_{i}$ its probability that
depends on the Na content, and $E_{i}$ its energy. Similarly $f$ labels the
final states. $\hbar \omega $ is the energy of the incoming light
represented by the vector potential $\mathbf{A}$.

In order to simulate the measured spectra, we have broadened the delta
functions appearing in Eq. (\ref{ixas}) replacing them by a Lorentzian line
shape with a full width of 0.5 eV at half maximum (FWHM) at the $L_{3}$%
--edge and 0.7 eV at the $L_{2}$--edge due to different life time broadening
effects.\newline

\section{Results}

\subsection{The structure of the ground state}

There are already indications that the Co ions in $\mathrm{Na}_{x}\mathrm{CoO%
}_{2}$ are in a low spin state \cite{Motohashi_PRB03, Lang_PRB05}, supported
also by the interpretation of XAS measurements using polarized light \cite%
{Wu_PRL05, Kroll_UJS10}. This means that crystal field effects dominate over
the exchange terms. In fact, from the parameters that best fit the
experiment (see Table \ref{best fit}), one sees that $10Dq=1.2$ eV. As
mentioned in the previous section, this corresponds to the ``ionic'' part of
the crystal field. The actual splitting
is much larger due to the covalency
effects which are also strong in the system. In table \ref{table1} we
present an overview of the configurations present in the ground state for
clusters with nominal oxidation state
$\mathrm{Co^{3+}}$ and $\mathrm{Co^{4+}}$.

\begin{table}[h]
\begin{center}
\begin{tabular}{c|cc||cc|c}
\multicolumn{2}{l}{$\mathrm{Co^{3+}}$} & \text{ \hspace{5mm}} & \text{
\hspace{5mm}} & \multicolumn{2}{l}{$\mathrm{Co^{4+}}$} \\
\multicolumn{3}{l||}{Groundstate} & \multicolumn{3}{r}{Groundstate} \\ \hline
$\mathrm{d^6}$ & 0.30 &  &  & $\mathrm{d^5}$ & 0.14 \\
$\mathrm{d^7L^1}$ & 0.47 &  &  & $\mathrm{d^6L^1}$ & 0.40 \\
$\mathrm{d^8L^2}$ & 0.20 &  &  & $\mathrm{d^7L^2}$ & 0.35 \\
$\mathrm{d^9L^3}$ & 0.03 &  &  & $\mathrm{d^8L^3}$ & 0.10 \\
$\mathrm{d^{10}L^4}$ & 0.00 &  &  & $\mathrm{d^9L^4}$ & 0.01 \\
&  &  &  & $\mathrm{d^{10}L^5}$ & 0.00 \\ \hline
\end{tabular}%
\end{center}
\caption{{\protect\small Overview of the hole distribution for the ground
states. The numbers give the hole distribution in percent.}}
\label{table1}
\end{table}

The ground state for $\mathrm{Co^{3+}}$ is a singlet with $A_{1g}$ symmetry
(for the sake of clarity we use capital letters to denote the irreducible
representations of many-body states). For the cluster representing nominal $%
\mathrm{Co^{3+}}$, the most important configurations in the ground state are
$d^{6}$ (30\%) and $d^{7}L^{1}$ (47\%), indicating a very strong covalency
in the system. Note that the $d^{6}$ population is smaller than the $d^{7}L$
population
and considerably less than 50\% of the total.
Such a behavior can only be explained if one takes more than one
ligand hole into account. In a configuration containing only $d^{6}$ and
$d^{7}L$ with a positive charge transfer gap $\Delta
_{CT}=E(d^{n+1}L)-E(d^{n})$ (see table III), %\ref{best_fit}),
the $d^{6}$ population will always be larger or equal to $d^{7}L$, but will
never be smaller. Things change if one includes also configurations
containing two ligand holes $d^{8}L^{2}$ with sufficiently large hopping
terms. In this case the relative hole population changes towards the $d^{7}L$
configuration, so that even for a positive charge transfer gap $\Delta _{CT}$
the dominant configuration can include a ligand hole.
This is
a rather important result since it demonstrates that the
hybridization between all those states involving one or more
ligand holes is large enough to produce a low energy state which
in fact is lower than that of the starting $d^{6}$ configuration
containing no ligand holes.
The $d^{6}$
configuration corresponds to $t_{2g}^{6}$ (all holes with $e_{g}$ symmetry
occupied). For simplicity, to reduce the size of the matrices in the fitting
procedure, we neglect the configurations $t_{2g}^{5}e_{g}^{1}$. This is
justified by the fact that the interaction term which mixes the 3d
configurations $t_{2g}^{6}$ and $t_{2g}^{5}e_{g}^{1}$ is $\lambda =\sqrt{3}%
(F_{2}-5F_{4})\simeq 0.3$ eV (see Table \ref{best fit}), is much smaller
than the effective crystal-field splitting $\sim 3$ eV. The amount of $%
t_{2g}^{5}e_{g}^{1}$ 3d configurations can be calculated by perturbation
theory, but in any case, its influence on the XAS spectra is of the order of
1\%. Similarly, for $\mathrm{Co^{4+}}$, we have neglected states with more
than one 3d $t_{2g}$ hole.

With these simplifications and because of the neglect of interactions
between electrons at the O sites, the number of relevant states in the
subspace in which the ground state for $\mathrm{Co^{3+}}$ is, reduced to 7
singlets with $A_{1g}$ symmetry. Similarly, the ground state for $\mathrm{%
Co^{4+}}$ is obtained from six identical 38x38 matrices and corresponds to a
spin doublet with $T_{2g}$ symmetry (this means three-fold orbital
degeneracy).
%When the $O_h$ symmetry is broken, the $E_{g}$ singlets mix with the ground state (see next subsection).

Due to the symmetry of the cluster ($O_{h}$), the 3d holes conserve its
symmetry when they hop to the O 2p states. Neglecting the small amount of 3d
$t_{2g}^{5}e_{g}^{1}$ states in $\mathrm{Co^{3+}}$, the only possibility for
a ligand hole is then a linear combination with $e_{g}$ symmetry giving the
configuration $d_{t_{2g}^{6}e_{g}^{1}}^{7}L_{e_{g}^{1}}^{1}$.

\begin{figure}[t]
\begin{center}
\subfigure[]{\label{configuration_L1}
\includegraphics[bb=0 0 312 159,width=0.8\columnwidth,angle=0,clip]{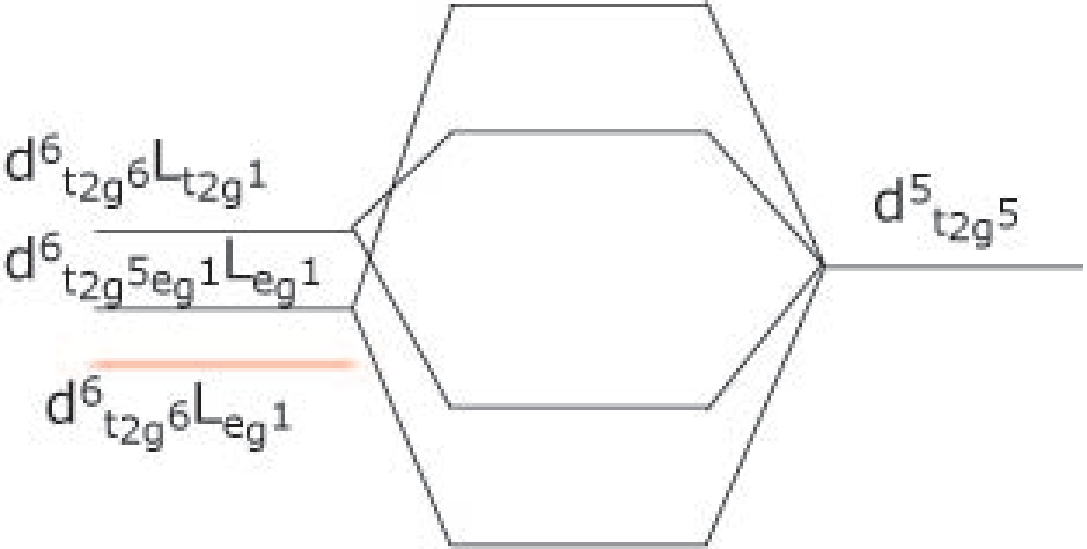}}
\subfigure[]{\label{configuration_L2a}
\includegraphics[width=0.8\columnwidth,angle=0,clip]{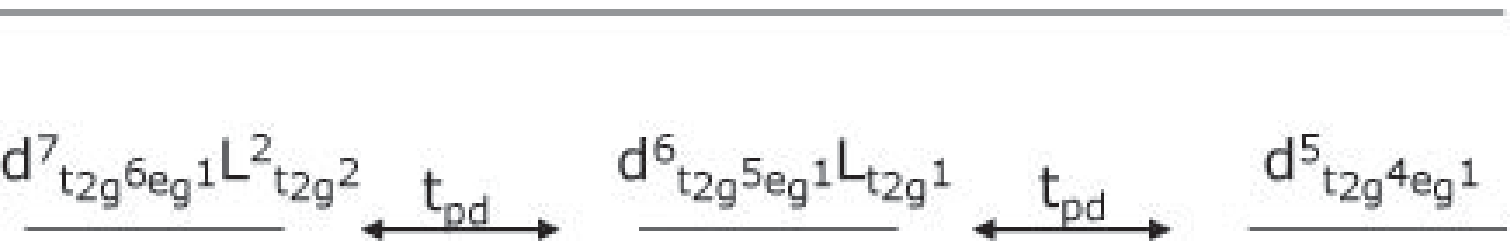}}
\subfigure[]{\label{configuration_L2b}
\includegraphics[width=0.8\columnwidth,angle=0,clip]{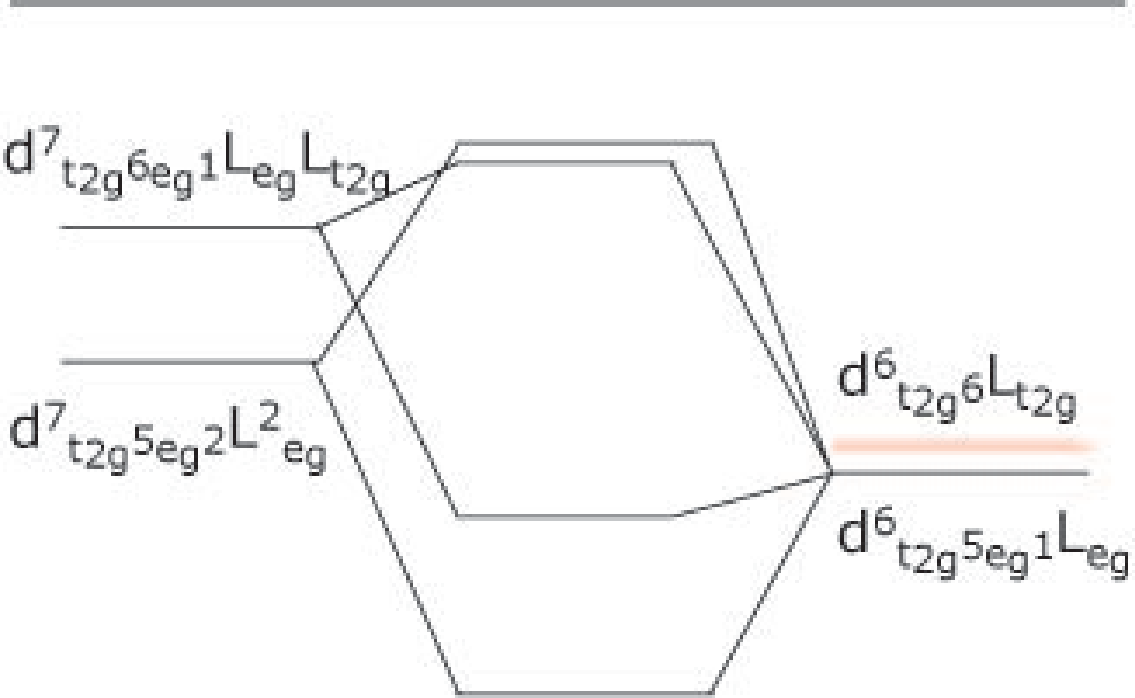}}
\end{center}
\caption{{\protect\small Sketches of the energy scheme for different hole
configurations of $Co^{4+}$. (a): Hybridization between states with none and
one ligand hole, (b) and (c): Hybridization between states with one and two
ligand holes.}}
\end{figure}
\begin{table}[h]
\begin{center}
\begin{tabular}{|c|c||c|c||c|c}
\multicolumn{6}{|l}{$\mathrm{Co^{4+}}$ (five holes)} \\
\multicolumn{6}{|l}{Ground state} \\ \hline
$\mathrm{d^5}$ & 0.14 & $\mathrm{d^6L^1_{e_g^1}}$ & 0.37 & $\mathrm{%
d^7L^2_{e_g^2}}$ & 0.28 \\
&  & $\mathrm{d^6L^1_{t_{2g}^1}}$ & 0.04 & $\mathrm{d^7L^2_{e_g^1t_{2g}^1}}$
& 0.07%
\end{tabular}%
\end{center}
\caption{{\protect\small Hole distribution for the ground state of the $%
\mathrm{Co^{4+}}$ cluster, including information of symmetry up to
configurations with two ligand holes. The numbers give the hole distribution
in percent.}}
\label{table2}
\end{table}

A more complicated picture is expected for the cluster corresponding to $%
\mathrm{Co^{4+}}$, since it contains five holes meaning that there will
always be one hole with $t_{2g}$ symmetry in the system (either at the Co
site or distributed among the O ions). For the parameters that fit the
experiment, the ground state of this cluster is a spin doublet with orbital
symmetry $T_{2g}$. This three-fold degeneracy in the point group $O_{h}$ is
split by the rhombohedral distortion as discussed in the next subsection.
While an overview of the structure of the ground state is given in table \ref%
{table1}, the distribution among different possible configurations up to two
ligand holes for the ground state of the $\mathrm{Co^{4+}}$ octahedra is
shown in table \ref{table2}. Note that states with one and two ligand holes
dominate the composition of the ground state. The $d^{5}$ configuration is
very simple. It consists of four 3$d_{e_{g}}$ holes and one 3$d_{t_{2g}}$
hole, or equivalently it consists of five 3$d_{t_{2g}}$ electrons ($%
d_{t_{2g}^{5}}^{5}$). The configurations with one ligand hole with the same
symmetry are $d_{t_{2g}^{5}e_{g}^{1}}^{6}L_{e_{g}^{1}}$, and $%
d_{t_{2g}^{6}}^{6}L_{t_{2g}^{1}}$. The former has a larger hybridization
with the $d^{5}$ configuration because the hopping between 3d and 2p $e_{g}$
or $t_{2g}$ electrons involves the Slater-Koster parameter $(pd\sigma )$ or $%
(pd\pi )$ respectively, and the former has a larger magnitude. For
configurations with two ligand holes there are two configurations with $%
T_{2g}$ symmetry: $d_{t_{2g}^{6}e_{g}^{1}}^{7}L_{e_{g}^{1}}L_{t_{2g}^{1}}$
and $d_{t_{2g}^{5}e_{g}^{2}}^{7}L_{e_{g}^{2}}$. They both hybridize with the
dominant one hole state $d_{t_{2g}^{5}e_{g}^{1}}^{6}L_{e_{g}^{1}}$. The only
difference is that the first configuration couples via $t_{t_{2g}}$ hopping
to the one hole state while the second configuration couples via a $%
t_{e_{g}} $ hopping which is larger and results in a stronger occupation of
this state (see table \ref{table2}). Although it is allowed by symmetry, the
above mentioned states with two ligand holes do not hybridize with the state
$d_{t_{2g}^{6}}^{6}L_{t_{2g}}$ due to the particular structure of the wave
functions.

>From the above described distribution of holes in the Co$^{3}$ and Co$^{4}$
clusters, it turns out that the real Co valence for each cluster is 2.04 and
2.56 respectively,
emphasizing again
the very strong degree of covalency in these systems.

\subsection{Breaking of the octahedral symmetry}

The edge-sharing $\mathrm{CoO_{6}}$ octahedra in a $\mathrm{CoO_{2}}$ layer
of $\mathrm{Na}_{x}\mathrm{CoO}_{2}$ are compressed along the $c$ axis
(oriented parallel to the (111) direction of one octahedra, see Fig. \ref%
{distortion}). This distortion reduces the point group symmetry of our
cluster to $D_{3d}$. As explained above, the ground state of the $\mathrm{%
Co^{4+}}$ cluster belongs to the $T_{2g}$ representation of $O_{h}$. This
representation is split in $D_{3d}$ as $T_{2g}=A_{1g}^{\prime
}+E_{g}^{\prime }$, where to avoid confusion, the representations of $D_{3d}$
are primed. Labelling as $|xy\rangle $, $|yz\rangle $, and $|zx\rangle $ the
basis functions of the representation $T_{2g}$ (which in our case correspond
to the many-body wave functions of the $\mathrm{Co^{4+}}$ cluster), the
basis functions that transform irreducibly in $D_{3d}$ can be written as
\begin{equation}
|A_{1g}^{\prime }\rangle =\frac{1}{\sqrt{3}}(|xy\rangle +|yz\rangle
+|zx\rangle )  \label{g1}
\end{equation}
for the state invariant under operations in $D_{3d}$ and
\begin{equation}
|E_{g}^{\prime }\pm \rangle =\frac{1}{\sqrt{3}}(|xy\rangle +e^{\pm i\frac{%
2\pi }{3}}|yz\rangle +e^{\pm i\frac{4\pi }{3}}|zx\rangle )
\end{equation}
for the doubly degenerate states that transform like $E_{g}^{\prime }$. A 3d
orbital with the symmetry $A_{1g}^{\prime}$ looks like a $3z^{2}-r^{2}$
orbital and points into the direction of the crystallographic $c$ axis. In a
one-electron picture, for an adequate ordering of the energy of the
orbitals, the $\mathrm{Co^{4+}}$ cluster would correspond to $e_{g}^{\prime
} $ orbitals completely filled with 4 electrons and a half filled $%
a_{1g}^{\prime }$ orbital \cite{Bayrakci_PRB04}.

\begin{figure}[t]
\begin{center}
\subfigure[]{\label{distortion}
\includegraphics[width=0.45\columnwidth,angle=0,clip]{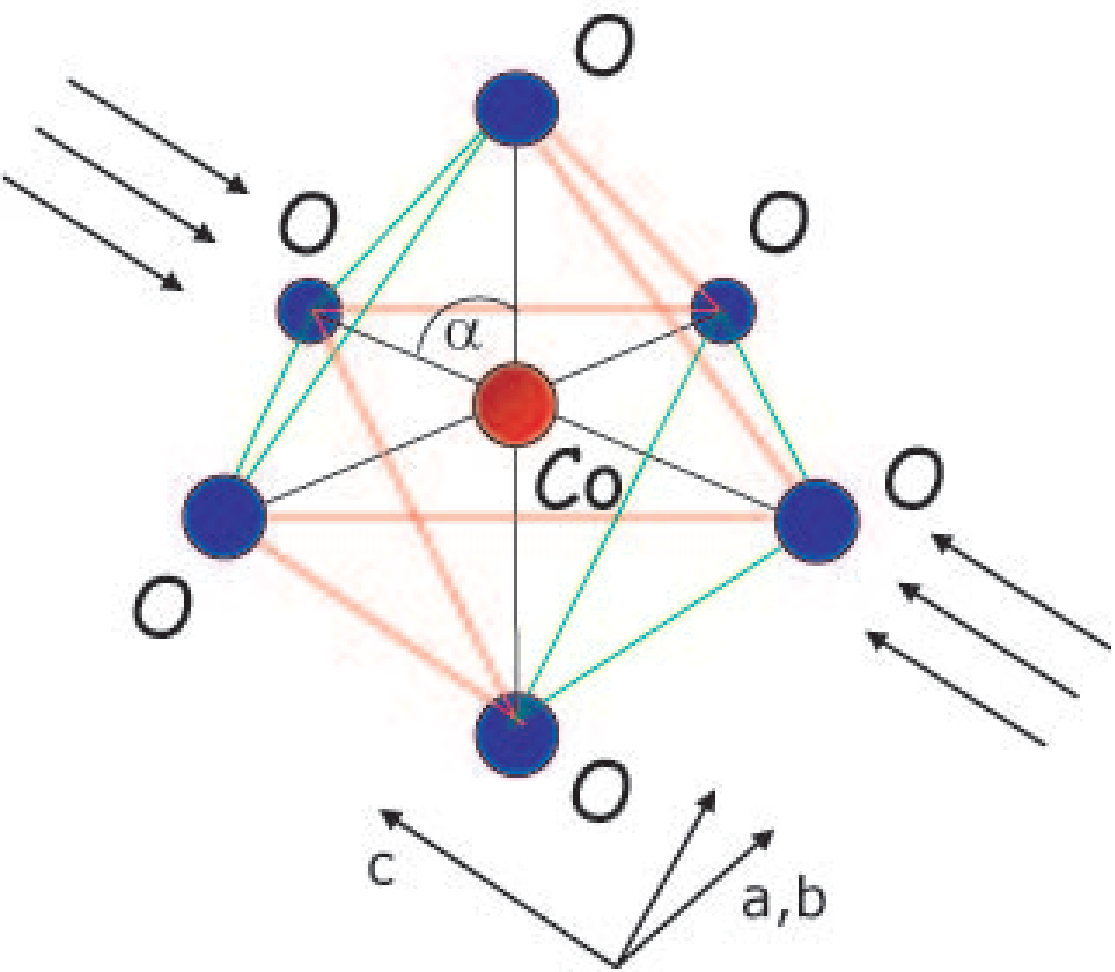}} %
\subfigure[]{\label{D_trg}\includegraphics[width=0.45%
\columnwidth,angle=0,clip]{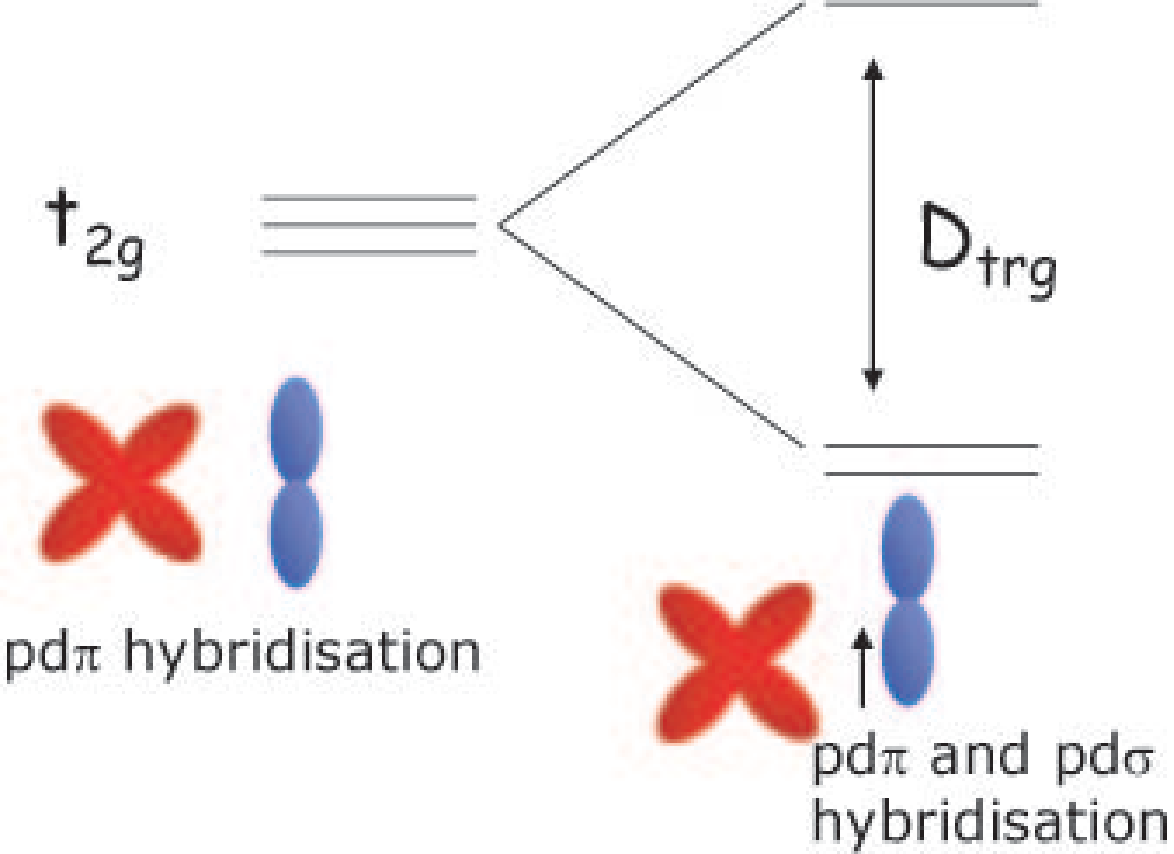}}
\end{center}
\caption{{\protect\small (a): Distortion of the octahedra. (b): Splitting of
the $t_{2g}$ states in a trigonal crystal field.}}
\end{figure}

The rhombohedral distortion of the $\mathrm{CoO_{6}}$ octahedra is estimated
by the deviation of the Co-O-Co bond angle from $\mathrm{90^{\circ }}$. For
example for $\mathrm{x\approx 0.7}$ Huang \textit{et al.} find an angle $%
\alpha \approx 96^{\circ }$ \cite{Huang_PRB04} which leads to a change of
the thickness of one octahedra by 10\%, assuming that the Co-O bond distance
remains constant during the distortion.

>From the fit of unpolarized XAS spectra \cite{Wu_PRL05}, Wu et al.
estimated
$D_{trg}=1$ eV for the splitting $D_{trg}=\langle E_{g}^{\prime }\pm |%
\mathcal{H}|E_{g}^{\prime }\pm \rangle -\langle A_{1g}^{\prime }|\mathcal{H}%
|A_{1g}^{\prime }\rangle $ between the many-body ground state
$A_{1g}^{\prime }$ and the excited states $|E_{g}^{\prime }\pm
\rangle $. Instead, Koshibae and Maekawa using a point charge
model obtained a negligigle splitting and suggested that the band
splitting at the zone center obtained in band structure
calculations is dominated by the effective Co-Co hopping mediated
by intermediate O atoms \cite{Koshibae_PRL03}.
However, one expects that the splitting is dominated by covalency
effects and that the ionic contribution plays a minor role.
Very recently
$D_{trg}=0.315$eV has been obtained by an ab-initio quantum
chemical configuration interaction method in a CoO$_{6}$ cluster
\cite{ll}.
This is the most reliable result. The method used is known to provide
accurate results for charge conserving excitations and has been used
succesfully in a large family of strongly correlated systems \cite{ll}.

To check if the distortion affects our results for cubic
symmetry, we have estimated the matrix element
$Q_{mix}=\langle E_{g}^{\prime }\pm |
\mathcal{H}|E_{g}^{\prime }\pm ,2\rangle $ between the $E_{g}^{\prime }$
states and the lowest exited states $|E_{g}^{\prime }\pm ,2\rangle $ of the
same symmetry. These excited states belong to the irreducible representation
$E_{g}$ of $O_{h}$ and therefore do not hybridize with the ground state in
cubic symmetry. The only part of the Hamiltonian where the trigonal
distortion comes into play is a change in the hopping part through
changes in relative orientations of the orbitals, and O-O distances.
We assumed an $r^{-3}$ distance dependence for the
hopping between $p$ orbitals \cite{Harrison}.
We find $Q_{mix}=-0.09$ eV.
Since $|Q_{mix}|$ is very small compared to the other energies of the model, in
particular $10Dq$ and the separation between $T_{2g}$ and $E_{g}$ levels, we
can safely use the states obtained in cubic symmetry for the calculation of
the XAS spectra.

A ground state with $E_{g}^{\prime }$ symmetry is incompatible with the XAS
data. Therefore only a positive $D_{trg}$ is consistent with experiment.
The value $D_{trg}=0.315$eV is smaller but of the order of magnitude of
the band width reported in ab-initio calculations.
A small value of $D_{trg}$
would be
consistent with the proposal of Koshibae and
Maekawa \cite{Koshibae_PRL03} that the energy splitting does not originate
in the crystal field, but is determined by the kinetic energy of the
electrons. Within this picture, translated to our many-body states, and
neglecting $D_{trg}$, the ground state for one $\mathrm{Co^{4+}}$ cluster
moving in a lattice of $\mathrm{Co^{3+}}$ is a Bloch state with total wave
vector $K=0$ in an effective Kagom\'{e} sublattice with point symmetry
$A_{1g}^{^{\prime }}$. To keep it simple, restricting the wave function to
only three sites, and one of the four Kagom\'{e} sublattices, the ground
state would be

\begin{equation}
|g\rangle =\frac{1}{\sqrt{3}}(|1,xy\rangle +|2,yz\rangle +|3,zx\rangle ),
\label{g3}
\end{equation}
where here $|i,\alpha \rangle $ denotes a state composed of the ground state
of one $\mathrm{Co^{4+}}$ cluster with symmetry $\alpha $ at site $i$, and
the (invariant $A_{1g}$) ground state of the $\mathrm{Co^{3+}}$ cluster at
the other two sites. However, the effective tight-binding Hamiltonian
considering only the most important hopping term and neglecting $D_{trg}$,
has a U(4)\ symmetry due to the equivalence between the four Kagom\'{e}
sublattices, and the different possibilities of breaking of this symmetry
leads to several scenarios for the ground state \cite{Indergand_PRB05}.

The difference between Eqs. (\ref{g1}) and (\ref{g3}) for the
polarization dependence of the intensity in the XAS spectra, can
be easily illustrated for the case in which the final states
$|f_{\alpha }\rangle $ in Eq. (\ref{ixas}) for a
$\mathrm{Co^{4+}}$ cluster have a core hole $2p_{\alpha }$
($\alpha =x$, $y$, or $z$) with given spin, and the rest of the
state (the holes of the 3d shell) belongs to the $A_{1g}$
irreducible representation. This intensity with a factor two
coming from spin, corresponds to the sum of the intensities of the
peak A in Fig. 3 (for a total momentum $j=3/2$ of the core hole)
and the corresponding peak at higher energies and lower intensity
for $j=1/2$ (A' in Fig.3). For incoming light in the direction
$\beta $, $\mathcal{H}_{L}$ is proportional to the momentum
operator in this direction $\hat{p}_{\beta }$. Then by symmetry,
all matrix elements $\langle f|\mathcal{H}_{L}|i\rangle $ entering
Eq. (\ref{ixas}) can
be related to $\gamma =\langle f_{x}|\hat{p}_{y}|xy\rangle $. Using Eq. (\ref%
{g1}), it is easy to see that the contribution of these states (summed over $%
\alpha $) to the intensity in the one-cluster picture, for \textbf{A} is
parallel and perpendicular to the $c$ axis becomes

\begin{equation}
I_{\parallel }^{1}=4|\gamma |^{2}/3\text{, }I_{\perp }^{1}=|\gamma |^{2}/3.
\label{i1}
\end{equation}
In the extended picture, naturally, the light operator is a sum over all
sites of that already described. Taking again only three sites for
simplicity, there are three possible final states for each core hole orbital
2p$_{\alpha }$ $|i,f_{\alpha }\rangle $ depending on which site $i$ was
excited by the core hole. Adding these contribution one finds, using Eq. (%
\ref{ixas})

\begin{equation}
I_{\parallel }^{ext}=I_{\perp }^{ext}=2|\gamma |^{2}/3.  \label{iext}
\end{equation}
Thus due to the loss of coherence, there is a factor 2 less
intensity for \textbf{A }parallel to\textbf{\ }$c$ in the XAS
spectra for the sum of the contributions of peaks A and A' (see
Fig. 3) in the case in which the hole is distributed over one
Kagom\'{e} sublattice. This factor is large enough to be detected
by the experiment. In the latter case, the result is the same as
for a ground state composed of statistical average of the three
$T_{2g}$ states and has no polarization dependence. The situation
is different if the breaking of the above mentioned U(4)\ symmetry
leads to some coherence between $T_{2g}$ states at the same site.
Note that the intensity is redistributed among the different
directions but keeps the same angular average. This is due to the
local character of the transition and the fact that only one
irreducible representation involved in the initial state, final
state and operator.

We shall show that our results point to an intermediate situation, but with
dominance of the lattice effects over the on-site distortion.

\subsection{The XAS intensity}

\begin{figure}[tbp]
\begin{center}
\includegraphics[width=1.0\columnwidth,angle=0,clip]{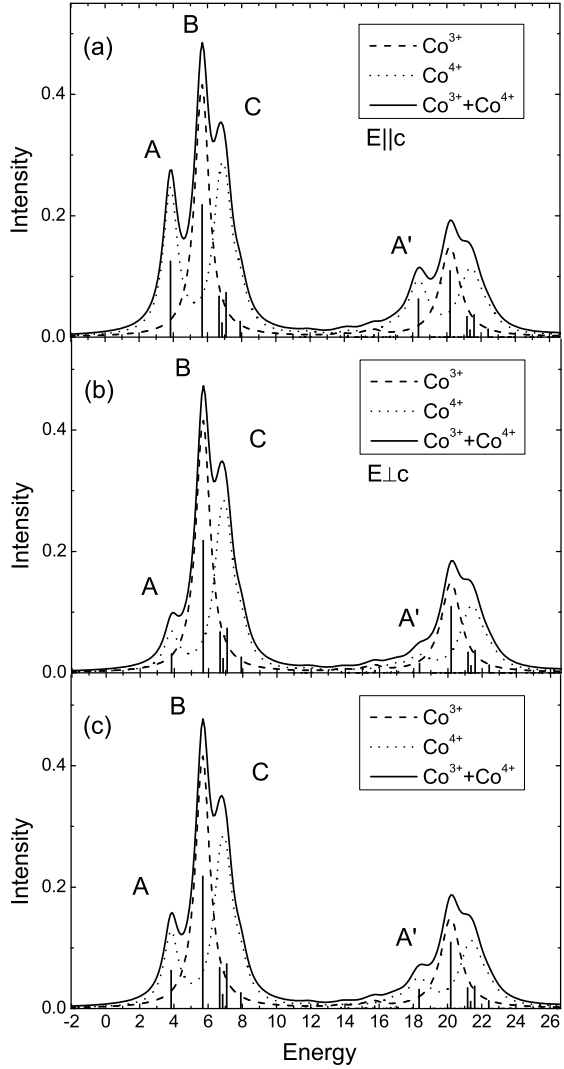}
\end{center}
\caption{{\protect\small XAS spectra for different polarizations: (a) $E||c$
and (b) $E\perp c$ Dashed line: $\mathrm{Co^{3+}}$ cluster. Dotted line: $%
\mathrm{Co^{4+}}$ cluster. Straight line: sum of both representing ($\mathrm{%
Na_{0.5}CoO_{2}}$). In (a) and (b) Eq. (\protect\ref{g1}) was used
for the ground state of the $\mathrm{Co^{4+}}$ cluster was used,
while (c) corresponds to Eq. (\protect\ref{g1}) giving a
polarization independent result.}} \label{bild}
\end{figure}

\begin{figure*}[t]
\begin{center}
\includegraphics[width=1.0\columnwidth,angle=0,clip]{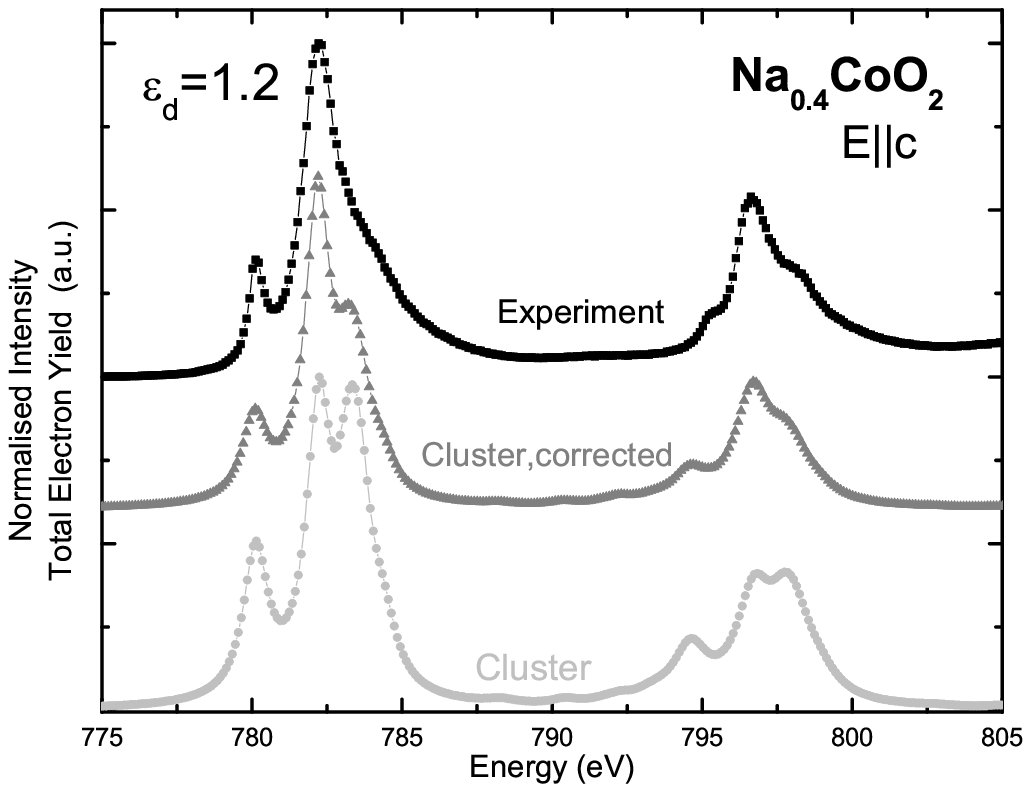} %
\includegraphics[width=0.98\columnwidth,angle=0,clip]{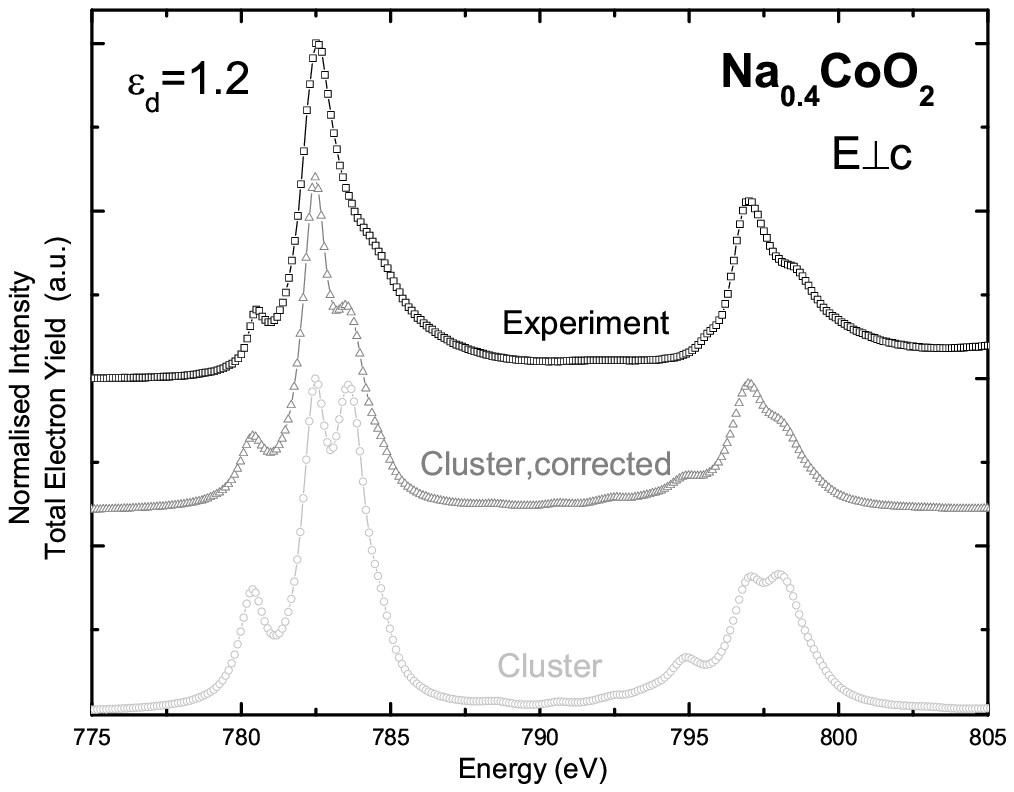} %
\includegraphics[width=1.0\columnwidth,angle=0,clip]{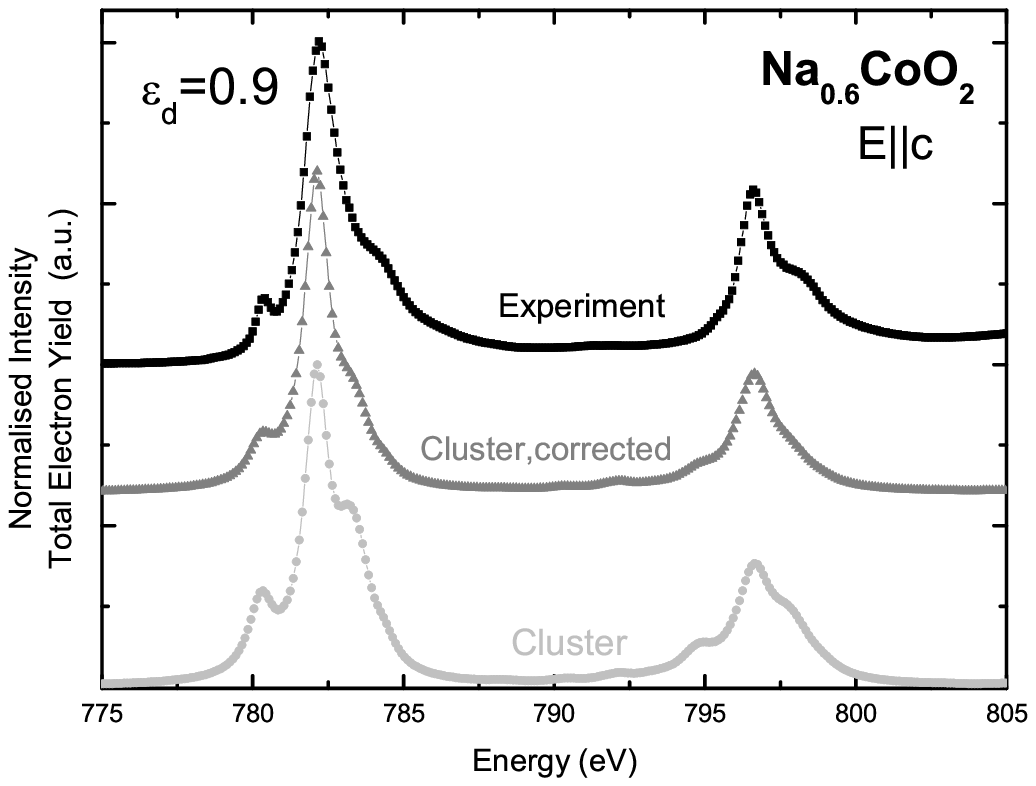} %
\includegraphics[width=1.0\columnwidth,angle=0,clip]{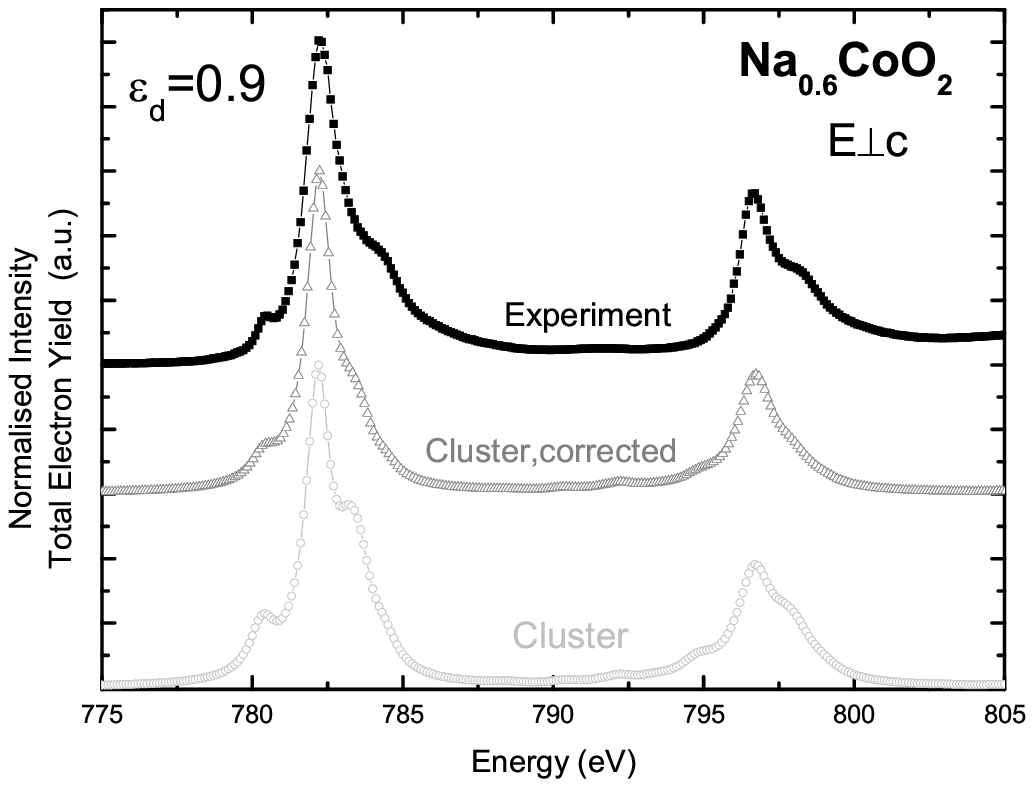}
\end{center}
\caption{{\protect\small Comparison of the theoretical results (bottom) with
experimental data (top) for x=0.4 and x=0.6 for polarizations indicated
inside each figure. In the middle curve the results are manipulated in such
a way, that the intensity of the Co$^{4+}$ cluster is decreased by a factor
2.0.}}
\label{comparison}
\end{figure*}

As mentioned above, the ground state for a $\mathrm{Co^{3+}}$ cluster is
invariant under the symmetry operations of $O_{h}$ and as a consequence, the
corresponding XAS intensity is independent of the polarization of the
incident light. Also, for $\mathrm{Co^{4+}}$, the XAS intensity for a
statistical average of the three states of the irreducible representation $%
T_{2g}$ ($|xy\rangle $, $|yz\rangle $ and $|zx\rangle $) (which would
correspond to a ground state like Eq. (\ref{g3})) leads to the same
intensity for \textbf{A}$||c$ and \textbf{A}$\perp c$.

Instead, using Eq. (\ref{g1}) for the ground state of
$\mathrm{Co^{4+}}$, one obtains a polarization dependence similar to the
experiment, but more pronounced: The ratio of the intensities for
excitations to a state with $A_{1g}$ symmetry plus a core hole is
$I_{||}^{1}/I_{\perp }^{1}=4$
(see Eq. (\ref{i1})) whereas the experimental ratio has been found to be
$I_{||}^{exp}/I_{\perp }^{exp}\approx 1.8$. Therefore the experiment suggests
an intermediate situation between the ground states (\ref{g1}) and (\ref{g3}%
). The actual many-body ground state of the system is out of the scope of
our cluster calculations. In order to simulate this situation, we have
calculated the spectrum as a weighted average of Eqs. (\ref{i1}) and
(\ref{iext}):\newline

\begin{equation}
I_{total}=(1-\xi )I_{f}^{1}+\xi I_{f}^{ext}  \label{itotal}
\end{equation}%
with $0\leq \xi \leq 1$. In order to get the observed experimental ratio for
the $A_{1g}$ peak $\xi $ has to be approximately $\xi =0.73$.

\begin{table}[tbp]
\begin{center}
\begin{tabular}{l|l|l}
$F_0=3.5$ & $pd\sigma=2.35$ & $10Dq^i=1.2$ \\
$F_2=0.2$ & $pd\pi=-1.0$ & $\epsilon_p=13$ \\
$F_4=0.006$ & $pp\sigma=0.8$ & $\Delta_{CT}(\mathrm{Co^{3+}})=2.9$ \\
$U_{dd}=4.52$ & $pp\pi=-0.2$ & $\Delta_{CT}(\mathrm{Co^{4+}})=-1.1$ \\
&  &
\end{tabular}%
\end{center}
\caption{{\protect\small Best fit parameters, all parameters given in eV.}}
\label{best fit}
\end{table}

In Fig. \ref{bild} we show the results for the XAS intensity for the
parameters indicated in Table \ref{best fit} that best fit the experiment.
As adjustable parameters, we have used $F_{0}$ (or $%
U_{d}=F_{0}+4F_{2}+36F_{4}$), $\epsilon _{\mathrm{O}}$ (or $\Delta_{CT} (%
\mathrm{Co}^{n+})= E(d^{n+1}L)-E(d^n)$), $10Dq$ ,($pd\sigma $), and ($%
pp\sigma $). We have taken $U_{dc}=U_{d}+0.15$ eV. $F_{2}$ and $F_{4}$ were
taken from values that fit approximately the energy separation between
different terms in atomic spectra \cite{moore}. We assumed the relations $%
(pd\pi)=-\sqrt{3}(pd\sigma )/4 $, and $(pp\pi )=-0.81(pd\sigma)/3.24$ \cite%
{Harrison}.

In terms of more fundamental parameters one has:

\begin{eqnarray}
\Delta (\mathrm{Co}^{3+}) & = & \epsilon_{\mathrm{O}} - \epsilon_{e_{g}}-3
F_0 +8 F_2 -33F_4 -pp\sigma +pp\pi  \nonumber \\
\Delta (\mathrm{Co}^{4+}) & = & \epsilon_{\mathrm{O}} - \epsilon_{e_{g}}-4
F_0 + 4 F_2 -37F_4/2 -pp\sigma +pp\pi,  \nonumber \\
\end{eqnarray}
and correspond to the minimum energy necessary to promote a Co 3d hole to a
linear combination of O 2p states in absence of Co-O hopping. If the O 2p
hole is localized at one site, $pp\sigma$ and $pp\pi$ are absent in these
equations and both $\Delta (\mathrm{Co}^{3+})$ increase in 1.0 eV.

To fit the experiments and explain the different observed structures,
several configurations with considerable amount of ligand holes (described
above) should be present in the ground state. This implies large covalency,
and this in turn points to high hybridization, a low charge transfer gap $%
\Delta_{CT}$ and low $U_{d}$ in comparison with other transition metal
oxides. Also, the relative position of the structures for $\mathrm{Co^{3+}}$
and $\mathrm{Co^{4+}}$ imply relatively large $10Dq$ and low spin
configurations. This is a rough explanation of the resulting parameters in
Table \ref{best fit}.

The $\mathrm{Co^{3+}}$ cluster (dashed line in Fig. \ref{bild}) shows only
one main peak at the $L_{3,2}$ edge (peak (B) in the figure \ref{bild})
which results from an excitation from the ground state with $A_{1g}$
symmetry to an excited state with $E_{g}$ symmetry and is polarization
independent. Comparing to the similar system $\mathrm{LiCoO_{2}}$ which
nominally contains only octahedra with four holes (i.e. a $\mathrm{Co^{3+}}$%
--central ion) \cite{vanElp_PRB91}, one observes a weak structure on the
high energy side of the $e_{g}$ peak which is missing here because of the
missing 2$p$--3$d$ core hole exchange interaction. The $\mathrm{Co^{4+}}$
cluster shows two main structures at both edges. The lower energy peak
structure (A) contains of a single peak and results from excitations into
states with $A_{1g} $ symmetry, while the higher energy peak structure (C)
originates from many excitations of low intensity to final states with $%
T_{1g}$ and $T_{2g}$ symmetry. Only the former peak (A) shows a polarization
dependence and is strong for incoming light polarized along the (111)
direction (crystallographic $c$ axis) and weak for polarization
perpendicular to the (111) direction (in-plane polarization). \newline

In Fig. \ref{comparison} we compare the theoretical results with
experimental data for the stoichiometries $\mathrm{Na_{0.4}CoO_{2}}$ and $%
\mathrm{Na_{0.6}CoO_{2}}$ as well as for polarization
perpendicular and parallel to the crystallographic $c$ axis. In
the theoretical curve at the bottom, we have used the weighted
average Eq. (\ref{itotal}) with $\xi =0.73$ which best fits the
polarization dependence of peaks A and A'. As explained above,
this average redistributes the spectral weight among the different
polarizations but does not affect the total intensity coming from
a Co$^{4+}$ cluster. As it is apparent comparing the bottom and
top curves in Fig. \ref{comparison}, the above mentioned average
is not enough to reach a quantitative agreement with experiment:
there is still a disagreement in the intensity ratio between that
coming from a Co$^{4+}$ cluster and that  originating from a
Co$^{3+}$ cluster. The theoretical contribution of  Co$^{4+}$
seems to be two times larger than in the experiment for x=0.4
(this ratio is a little bit less than two for x=0.6).
%While the redistribution of spectral weight caused by
%the partial loss of coherence improves the agreement with
%experiment, it does not affect the total intensity of the
%Co$^{4+}$ contribution.
Wet chemical redox analysis point to a proportion of Co$^{+4}$
smaller than that corresponding the nominal Na content
\cite{karp}. This can explain part of the discrepancy. Another
possible explanation for the discrepancy is the redistribution of
spectral weight due to itineracy of the final states. The effects
of the hopping in a model of two sites has been studied by G.A.
Sawatzky and A. Lenselink \cite{lense}. In our case, these effects
seem to be of minor  importance for the peak A, because the
expected magnitude of the hopping $\sim $ 0.1 eV
\cite{Koshibae_PRL03} is
much smaller than the energy difference between excited states involved ($%
\sim $ 1.9 eV). However they are certainly important for the
higher energy multiplets.

Note that our neglect of the exchange interaction between the core
hole and the 3d holes leads to slight change in the shape of the
curves, but does not affect the total intensities for both
clusters.

Spin-orbit coupling inside the 3d shell, which we have neglected might also
lead to a reduction of the intensity of the low energy peak at the $L_2$%
--edge that corresponds to peak A as an effect of selection rules \cite%
{Gawelda_JACS06}. However, in our case, the effect of spin-orbit coupling is
small compared to the intersite hopping terms $t \sim 0.1$ eV \cite%
{Koshibae_PRL03}. Therefore, our approach of considering a sixfold
degenerate ground state for Co$^4$ ($T_{eg}$ doublet) instead of split
doublet and quartet states is justified. Therefore, we expect that the
reduction of intensity due to spin-orbit coupling inside the 3d shell is
small.

The relative peak positions do not change upon doping, except for the $%
A_{1g} $ peak. The distance between the main ($E_{g}$) peak and the $A_{1g}$
peak becomes smaller with increasing $x$ (decreasing amount of $\mathrm{%
Co^{4+}}$ ions) \cite{Kroll_UJS10}. This change is best simulated by a
variation of the ionic crystal field $10Dq$ from 1.2 eV ($x$=0.4) to 0.9 eV (%
$x$=0.6). A decreasing value of $10Dq$ is also supported by Huang \textit{et
al.} who performed neutron diffraction on powder samples and found an
increasing Co--O bond length with increasing doping \cite{Huang_PRB04}. This
would change the total crystal field splitting (the ionic term $10Dq$ plus
the effects of hopping), but a slight change in the hopping parameters does
not change the relative peak positions.

\section{Summary and discussion}

We have investigated the local electronic structure around a Co atom in $%
\mathrm{Na}_{x}\mathrm{CoO}_{2}$, by solving exactly a
$\mathrm{CoO}_{6}$ cluster (a basic octahedron of the system)
containing the Co 3d and O 2p valence electrons, and with the
appropriate charge according to the Na doping. We have included
all interactions between 3d electrons in the cluster. For the
calculation of the polarization dependence of the XAS spectrum, we
have included the energy of the Co 2p core hole with its
spin-orbit coupling $\Delta _{SO}\sim $ 15 eV, and its repulsion
with the Co 3d holes. The exchange interaction between Co 2p and
3d holes has been neglected for simplicity. Due to comparatively
large magnitude of $\Delta _{SO}$, crystal-field effects and
covalency, one expect the influence of this term to be small in
comparison to lighter 3d transition metals \cite{Zaanen_PRB85}.

Within a purely local picture, assuming an $A_{1g}^{\prime}$ ground state
[Eq. (\ref{g1})] and disregarding for the moment the total intensity of the $%
\mathrm{Co^{4+}}$ contribution, we are able to reproduce the essential
features of the polarization dependent XAS experiments fairly well. Slight
differences at the shoulders of the largest peaks originate from the neglect
of the Co 2p-3d exchange interaction which would add an additional weak
structure in this region, as shown by comparison with the case $\mathrm{%
LiCoO_2}$ \cite{vanElp_PRB91}.

Quantitatively, a fully coherent $A_{1g}^{\prime }$ ground state,
as Eq. (\ref{g1}) leads to too large intensity ratio for peak A
(and A') of the Co$^{+4}$ contribution between light polarized
parallel or perpendicular to the trigonal axis. This points out
that a purely local picture is not fully consistent with
experiment. In fact, the best fit corresponds to a $\approx$ 70\%
loss of coherence produced by delocalization. This redistribution
of intensity among the different polarizations is consistent with
an itinerant picture based on four Kagom\'{e} sublattices hidden
in the $\mathrm{CoO_{2}}$ layer \cite{Koshibae_PRL03,
Indergand_PRB05} as a first approximation for the electronic
structure.

The above mentioned redistribution is not enough to explain the
large total amount of the Co$^{+4}$ contribution to the intensity.
Part of this discrepancy might be due to proportion of Co$^{+4}$
smaller than that suggested by the nominal Na content \cite{karp}.
A more quantitative calculation of the XAS spectrum requires to
take into account the itineracy of holes in the ground state and
the excited states after the
creation of the core hole, which leads to spectral weight transfer between Co%
$^{+4}$ and Co$^{+3}$ contributions \cite{lense}. This is in principle
possible if one combines the knowledge of the local transition matrix
elements with an effective Hamiltonian for the motion of the holes through
the lattice. This approach has been followed to calculate different optical
properties of the cuprates.\cite{hyb,sim2,fei2,raman,lema,erol} However, in
this system this task is more difficult and is beyond the scope of the
present paper.

\section*{Acknowledgments}

This investigation was supported by the DFG project KL 1824/2, by the
Deutscher Akademischer Austauschdienst (DAAD), by PICT 03-12742 of
ANPCyT, Argentina and Canadian funding agencies
NSERC, CFI, and CIAR.
A.A.A. is partially supported by CONICET.

\end{document}